\documentclass[twoside,12pt]{article}
\usepackage{amssymb, amsmath, amsfonts}

\newtheorem{lemma}{Lemma}
\newtheorem{definition}{Definition}

\newtheorem{theorem}{Theorem}

\begin{document}

\title{Hydrodynamic type integrable equations  on a segment and a half-line.}
\author{ Metin G{\" u}rses$^a$, Ismagil Habibullin\footnote{e-mail:
habibullin\_i@mail.rb.ru, (On leave from Ufa
Institute of Mathematics, Russian Academy of Science,
Chernyshevskii Str., 112, Ufa, 450077, Russia)}$\,\,^b$
  and Kostyantyn Zheltukhin$^c$ \\
{\small $^a$Department of Mathematics, Faculty of Sciences}\\
{\small Bilkent University, 06800 Ankara, Turkey}\\
{\small e-mail gurses@fen.bilkent.edu.tr}\\
{\small $^b$Department of Mathematics, Faculty of Sciences}\\
{\small Bilkent University, 06800 Ankara, Turkey}\\
{\small e-mail habib@fen.bilkent.edu.tr}\\
{\small $^c$Department of Mathematics, Faculty of Sciences}\\
{\small Middle East Technical University 06531 Ankara, Turkey}\\
{\small e-mail zheltukh@metu.edu.tr}\\}

\begin{titlepage}

\maketitle

\begin{abstract}
The concept of integrable boundary conditions is applied to
hydrodynamic type systems. Examples of such boundary conditions
for dispersionless Toda systems are obtained. The close relation
of integrable boundary conditions with integrable reductions of
multi-field systems is observed. The problem of consistency of
boundary conditions with the Hamiltonian formulation is discussed.
Examples of Hamiltonian integrable hydrodynamic type systems on a
segment and a semi-line are presented.
\end{abstract}

{\it Keywords:} hydrodynamic type equations, integrable boundary
conditions, symmetries, Toda systems, Hamiltonian representation.

\end{titlepage}

\section{ Introduction.}

The theory of the integrable hydrodynamic type of systems
\begin{equation}\label{hydrodynamicType}
u_t^i=v^i_j(u)u_x^j,\quad i,j=1,2,...N,
\end{equation}
was initiated by S.P.Novikov and  B.A.Dubrovin
\cite{DubrovinNovikov}, and S.P.Tsarev \cite{Tsarev}. Here in
equation (\ref{hydrodynamicType}) summation over the repeated
indices is assumed and $ u$ is an N-component column vector of the
form $ u=(u^1,u^2,...u^N)^t$. Such systems have a variety of
applications in gas dynamics, fluid mechanics \cite{b,km1,km2,gt},
chemical kinetics, Whitham averaging procedure
\cite{gk,zh1,Krichever,Flaschka}, differential geometry and
topological field theory. We refer to \cite{Ferapontov,p} for
further discussions and references.

In the present article, a problem of finding boundary conditions
for hydrodynamic type equations consistent with the integrability
property is studied for a special case of the system
(\ref{hydrodynamicType}) called dispersionless Toda lattices
(\cite{gz,fs,li}). Actually we assume that equation
(\ref{hydrodynamicType}) admits a Lax representation on the
algebra
\begin{equation}\label{Algebra}
A=\left\{\sum\limits_{-\infty}^\infty u_i(x)p^i\,:\, u_i \quad
\mbox{decay sufficiently rapidly as} \quad x\to \pm\infty \right\}
\end{equation}
with the following Poisson bracket
$$
\{f,g\}=p\left( \frac{\partial f}{\partial p}\frac{\partial g}{\partial x}-
\frac{\partial f}{\partial x}\frac{\partial g}{\partial p}\right)\qquad f,g\in A.
$$
Such equations, for example, appear in the fluid mechanics as reductions of Benny
moment equations \cite{b}-\cite{p}.

Our definition of consistency of boundary conditions with the
integrability (see \cite{ggh,h,aggh}) is based on the notion of
symmetries. A constraint of the form
\begin{equation}\label{BoundaryCondition}
f(t, u, u_{[1]},..., u_{[k]})|_{x=x_0}=0,
\end{equation}
where $u_{[j]}=\frac{\partial^j}{\partial x^j}u$, imposed at some
point $x_0$ is called a boundary condition at this point. Boundary
value problem (\ref{hydrodynamicType}), (\ref{BoundaryCondition})
(or simply boundary condition (\ref{BoundaryCondition})) is called
consistent with the symmetry
\begin{equation}\label{Symmetry}
u^i_{\tau}=\sigma(t,x, u, u_{[1]},..., u_{[m]})
\end{equation}
if (\ref{hydrodynamicType}) and (\ref{Symmetry}) are compatible
under the constraint  (\ref{BoundaryCondition}). More precisely,
we mean the following: Differentiation of
(\ref{BoundaryCondition}) with respect to $\tau$ yields
\begin{equation}\label{DifferentialConsequence}
\sum\frac{\partial f}{\partial u^i_{[n]}}(u^i_{[n]})_{\tau}=0,
\end{equation}
where $\tau$-derivatives are replaced by means of equation
(\ref{Symmetry}).
\begin{definition}.\label{Def_1}
Boundary value problem (\ref{hydrodynamicType}),
(\ref{BoundaryCondition}) is consistent with the symmetry
(\ref{Symmetry}) if (\ref{DifferentialConsequence}) holds
identically by means of (\ref{BoundaryCondition}) and its
differential consequences obtained by differentiating with respect
to $t$.
\end{definition}

Note that, since constraint (\ref{BoundaryCondition}) is valid
only for $x=x_0$ it cannot be differentiated with respect to $x$.
For this reason it is convenient to exclude the $x$-derivatives of
dependent variable $u$ from our scheme. By solving equation
(\ref{hydrodynamicType}) for $u^i_x$ one gets $ u_x=v^{-1} u_t,$
where $v^{-1}$ is the matrix inverse to $v_j^i( u)$. Similarly,
$u_{xx}=(v^{-1} u_t)_x=(v^{-1})_x u_t+v^{-1} u_{tt}$ is expressed
through $ u, u_t, u_{tt}$ and so on. As a result one can rewrite
boundary condition (\ref{BoundaryCondition}) and symmetry
(\ref{Symmetry}) taken at the point $x_0$ as
\begin{equation}\label{constraint}
f_1(t, u,  u_t,...)=0
\end{equation}
and
\begin{equation}\label{associated}
u_{\tau}=\sigma_1(t,x_0, u,  u_t,...).
\end{equation}
Now the consistency requirement can be reformulated as follows.
Boundary condition (\ref{BoundaryCondition}) is consistent with
(\ref{Symmetry}) if differential constraint (\ref{constraint}) is
consistent with the associated $\tau$-dynamics (\ref{associated}).
We call the boundary condition consistent with integrability if it
is consistent with an infinite dimensional subspace of symmetries.
Hydrodynamic type  system given in (\ref{hydrodynamicType})
defines an $N$-dimensional dynamical system and the boundary
condition (\ref{constraint}) defines an hypersurface in
$N$-dimensional space of functions. Thus, due to the remark above
an integrable boundary condition is closely connected with
reductions of the associated system (\ref{associated}) compatible
with integrability \cite{gz}. Below we use this important
observation in order to find symmetry consistent boundary
conditions.

Boundary conditions passed the symmetry test are then tested for
consistency with the conserved quantities, Hamiltonian structures
and the complete integrability property of system
(\ref{hydrodynamicType}). It is remarkable that some of the
boundary conditions satisfy also these additional requirements and
thus allow one to reduce (\ref{hydrodynamicType}) to a completely
integrable Hamiltonian system on a segment and a half-line.

The paper is organized as follows. In Section 2 some integrable
boundary conditions for Toda system are  derived and it is showen that
these boundary conditions are compatible with infinite number of
symmetries. The relation between the integrable reductions of
N-system and the integrable boundary conditions is considered in
Section 3. It is observed that some integrable boundary
conditions lead to integrable reductions. In section 4 we discuss
the compatibility of the integrable boundary conditions, found in
the previous sections, with the Hamiltonian formulation. We show
that some boundary conditions are indeed compatible with the
Hamiltonian formulation and also with an infinite class of
symmetries. In all sections up to Section 4 only $N=2$
systems are considered. In Section 5 we study  $N=3$ systems which give other
examples of hydrodynamic type equations. For this case, some integrable
boundary conditions compatible with infinite number of symmetries
and boundary conditions compatible with the Hamiltonian formulation are found.

\section{Integrable Boundary conditions for the
Toda system.}\label{sec_int_bound_cond}

In this section we study the well known example of integrable
model \cite{fs1}
\begin{equation}\label{Eqn_Toda1}
\begin{array}{l}
S_t=P_x,\\
P_t=PS_x\\
\end{array}
\end{equation}
called Toda system, admitting the Lax representation on the
algebra of Laurent series (\ref{Algebra})
\begin{equation}\label{Lax_Eqn_Toda1}
L_t=\{(L)_{\ge 0},L\},
\end{equation}
where
\begin{equation}\label{Lax_Funct_Eqn_Toda}
L=p+S+Pp^{-1}.
\end{equation}
The corresponding hierarchy of symmetries of the Toda system
(\ref{Eqn_Toda1}) is
\begin{equation}\label{Eqn_Toda_Hierarchy1}
L_{t_n}=\{L,(L^n)_{\ge 0}\}.
\end{equation}
Recursion operator corresponding to the above hierarchy is (for
calculation of recursion operator see  \cite{gz}, \cite{z})
\begin{equation}\label{Toda_Rec_Oper_SP}
\mathfrak{R}=\left(
\begin{array}{ll}
  S & 2+P_xD_x^{-1}\cdot P^{-1}\\
2P & S+S_xPD_x^{-1}\cdot P^{-1}
\end{array} \right).
\end{equation}
In some cases it is convenient to consider the Toda
system(\ref{Eqn_Toda1}) in other variables. We write the Lax
function (\ref{Lax_Funct_Eqn_Toda}) as $L=p^{-1}(p+u)(p+v)$  that
is
\begin{equation}\label{uv_variables}
S=u+v,\quad P=uv.
\end{equation}
Then the Toda system (\ref{Lax_Eqn_Toda1})  gives
\begin{equation}\label{Eqn_Toda2}
\begin{array}{l}
u_t=uv_x\\
v_t=vu_x.\\
\end{array}
\end{equation}
Let us find boundary conditions compatible with an infinite number
of symmetries from the hierarchy (\ref{Eqn_Toda_Hierarchy1}). As a
boundary we take $x=0$. First we find boundary conditions
compatible with the first symmetry of the
hierarchy~(\ref{Eqn_Toda_Hierarchy1}). Assume that the boundary
condition depends on $P$, $S$ and can be solved with respect to
$S$.  So the boundary condition can be written as
\begin{equation}\label{Bound_Cond_Solved}
S=F(P),\quad x=0.
\end{equation}

\begin{lemma}\label{lemma1}
On the boundary $x=0$, the boundary condition of the form
(\ref{Bound_Cond_Solved}) compatible with the first symmetry of
the hierarchy (\ref{Eqn_Toda_Hierarchy1})
\begin{equation}\label{Symmetry_1}
\begin{array}{l}
S_{t_1}=2SP_x+2PS_x\\
P_{t_1}=2PP_x+2SPS_x
\end{array}
\end{equation}
is given by
\begin{equation}\label{Bound_Cond_Exact}
P=\displaystyle{\frac{(S+c)^2}{4}},\qquad x=0,
\end{equation}
\end{lemma}
{\bf Proof.} The boundary condition
(\ref{Bound_Cond_Solved}) is compatible with the symmetry
(\ref{Symmetry_1}) if on the boundary $x=0$
\begin{equation}\label{l1eqn1}
S_{t_1}=F'(P)P_{t_1}
\end{equation}
for all solutions of the equation (\ref{Eqn_Toda1}). Let us  find functions
$F$ for which the above equality holds. We rewrite the symmetry
(\ref{Symmetry_1}) in terms of  variables $S$, $P$ and their $t$
derivatives using the equation (\ref{Eqn_Toda1})
\begin{equation}
\begin{array}{l}
S_{t_1}=2SS_t+2P_t\\
P_{t_1}=2PS_t+2SP_t.
\end{array}
\end{equation}
Then we substitute $S_{t_1}$ and $P_{t_1}$ in (\ref{l1eqn1}), so
\begin{equation}
2SS_t+2P_t=F'(P)(2PS_t+2SP_t)
\end{equation}
From (\ref{Bound_Cond_Solved}) it follows that $S_t=F'(P)P_t$, so
\begin{equation}
2SF'(P)P_t+2P_t=F'(P)(2PF'(P)P_t+2SP_t).
\end{equation}
Hence
\begin{equation}
{F'}^2(P)=\displaystyle{\frac{1}{P}}.
\end{equation}
The above equation has a solution (\ref{Bound_Cond_Exact}). $\Box$

It is convenient to write the boundary condition
(\ref{Bound_Cond_Exact}) as
\begin{equation}\label{Bound_Cond_Transf}
P=\frac{S^2}{4},\quad x=0.
\end{equation}
by  shifting  $S$, the Toda system(\ref{Eqn_Toda1}) is invariant with respect to such shift.

\begin{lemma}\label{lemma2}
All the symmetries of the hierarchy (\ref{Eqn_Toda_Hierarchy1}) are
compatible with the boundary condition (\ref{Bound_Cond_Transf}).
\end{lemma}
{\bf Proof.}
The boundary condition (\ref{Bound_Cond_Transf}) is compatible with an
evolution symmetry
\begin{equation}\label{s1l2e2}
\left(
\begin{array}{l}
S\\
P
\end{array}
\right)_\tau=
\left(
\begin{array}{l}
\sigma\\
\pi
\end{array}
\right)
\end{equation}
if $\pi=\frac{1}{2}S\sigma$ for $P=\frac{1}{4}S^2$. That is under
the constraint (\ref{Bound_Cond_Transf}) the   symmetry
(\ref{s1l2e2}) should take the form
\begin{equation}\label{s1l2e1}
\left(
\begin{array}{l}
S_\tau\\
\frac{1}{2}SS_\tau
\end{array}
\right)=
\left(
\begin{array}{l}
\sigma\\
\frac{1}{2}S\sigma
\end{array}
\right)
\end{equation}
Evidently the first symmetry of the hierarchy
(\ref{Eqn_Toda_Hierarchy1}) has such a form. Let us show that the
recursion operator (\ref{Toda_Rec_Oper_SP}) preserves the property
(\ref{s1l2e1}). On the boundary $x=0$ we rewrite the recursion
operator (\ref{Toda_Rec_Oper_SP}) in terms of $t$ derivatives
using the Toda system(\ref{Eqn_Toda1}) as follows
\begin{equation}\label{Toda_Rec_Oper_SP_t}
\mathfrak{R}=\left(
\begin{array}{ll}
S+S_tD^{-1}_t & 2\\
2P+P_tD^{-1}_t & S\\
\end{array}
\right).
\end{equation}
Applying the recursion operator (\ref{Toda_Rec_Oper_SP_t}) to a symmetry (\ref{s1l2e1}) we obtaine a symmetry
\begin{equation}
\left(
\begin{array}{l}
S\\
P
\end{array}
\right)_{\tilde\tau}=
\left(
\begin{array}{l}
\tilde\sigma\\
\frac{1}{2}S\tilde\sigma
\end{array}
\right).
\end{equation}
$\Box$

\bigskip

We  also have the following boundary condition compatible with the
hierarchy~(\ref{Eqn_Toda_Hierarchy1}).
\begin{lemma}\label{lemma3}
On the boundary $x=0$, the boundary condition
\begin{equation}\label{Bound_Cond_P=0}
P=0
\end{equation}
is compatible with all symmetries of
the hierarchy (\ref{Eqn_Toda_Hierarchy1})
\end{lemma}
The above lemma is proved in the same way as lemma (\ref{lemma2}).

\bigskip

Another  boundary condition  comes from considering odd and even
solutions of the Toda system(\ref{Eqn_Toda1}). This boundary
condition is not compatible with all symmetries of the hierarchy
(\ref{Eqn_Toda_Hierarchy1}) but only with even ones.
\begin{lemma}\label{lemma4}
On the boundary $x=0$, the boundary condition
\begin{equation}
S=0
\end{equation}
is compatible with all even numbered symmetries of
the hierarchy~(\ref{Eqn_Toda_Hierarchy1}).
\end{lemma}
The above lemma is proved in the same way as lemma (\ref{lemma2}) using the square
of the recursion operator (\ref{Toda_Rec_Oper_SP_t}).

\section{Integrable reductions.}\label{Sec_Integr_Red}

Let us consider other equations admitting a Lax representation on
the algebra (\ref{Algebra}). For  a Lax function
$L=p^{-1}(p-u_{N})(p-u_{N-1})\dots(p-u_1)$, where $N>2$, we consider
the Lax equation
\begin{equation}\label{lax_eqn_N}
L_t=\{L,(L)_{\ge 0}\}.
\end{equation}
and an infinite hierarchy of symmetries
\begin{equation}
L_{t_n}=\{L,(L^n)_{\ge 0}\}\qquad n=1,2,\dots \, .
\end{equation}
For such equations we can not directly find boundary conditions
compatible with symmetries (see  section (\ref{sec_3dim_eqn})). So
we  use integrable reductions \cite{gz}.
\begin{definition}
A reduction of an integrable equation is called integrable if
reduced equation  is also integrable. That is the reduced equation
admits an infinite hierarchy of symmetries.
\end{definition}
In \cite{gz} it was shown that the following  reductions
\begin{equation}\label{reductions}
\begin{array}{l}
u_N=u_{N-1}=\dots =u_i=0, \quad i\ge 2,\\
u_N=u_{N-1}=\dots =u_j, \quad j\ge 1
\end{array}
\end{equation}
of the above equations are integrable. We note that for these
reductions the symmetries of the reduced equation are obtained by
the reduction of the symmetries of the original system.

If we have an integrable reduction such that symmetries of the
reduced system are obtained by the reduction of the symmetries of
the original system then  the  reduction can be taken as
integrable boundary conditions. Indeed, the original system is
invariant under the hierarchy of symmetries  and the reduced
system is invariant under the symmetries. Since reduction can be recovered from original system
and the reduced system it is also invariant under the
symmetries. So, taking the reductions (\ref{reductions}) as boundary conditions we
obtain symmetry invariant boundary conditions.
\begin{theorem}
For a system (\ref{lax_eqn_N}) the boundary conditions
$(u_N=u_{N-1}=\dots =u_i)|_{x=a}=0$, or $(u_N=u_{N-1}=\dots
=u_j)|_{x=a}$  (taking $x=a$ as the boundary) are
integrable.
\end{theorem}
Let us  take boundary conditions obtained in section
\ref{sec_int_bound_cond}. The condition $(P=\frac{S^2}{4})|_{x=0}$
in $u,v$ variables (\ref{uv_variables}) is $(u-v)|_{x=0}=0$. It
corresponds to a reduction $u=v$. The condition $P|_{x=0}=0$  in
$u,v$ variables is $(uv)|_{x=0}=0$. It corresponds to a reduction
$u=0$ (or $v=0$). The condition $S|_{x=0}=0$  in $u,v$ variables
is $(u+v)|_{x=0}=0$ . It does not correspond to reductions
considered above.

{\bf Remark.} If we take a reduction mentioned above as a boundary
condition then we can consider the corresponding reduced system.
Solutions of the reduced system obviously satisfy the main system
equations and  the boundary condition. For Toda system the
reduction $P=0$ leads to the equation
\begin{equation}
S_t=0.
\end{equation}
Its solution $S=f(x)$, for any differentiable function $f$,  gives
the solution of Toda system (\ref{Eqn_Toda1}) satisfying the
corresponding boundary condition (\ref{Bound_Cond_P=0}). The
reduction $P=\displaystyle{\frac{S^2}{4}}$ leads to the Hopf
equation
\begin{equation}
S_t=\frac{1}{2}SS_x.
\end{equation}
Its solution $S=h(2x+tS)$ gives the solution of Toda system
(\ref{Eqn_Toda1}) satisfying the corresponding boundary condition.
Here $h$ is any differentiable function of $x$ and $t$. To find a
solution of $N-$ system satisfying the integrable boundary
condition the method described above is very effective. We take
the corresponding reduction and the corresponding reduced ($N-1$)
system. Solving the reduced system gives automatically the
solution of the $N-$ system satisfying the integrable boundary
condition.

\section{Hamiltonian representation of the integrable boundary value problems.}

To obtain the Hamiltonian formulation of the Toda system
(\ref{Eqn_Toda1}) we use its Lax representation on the algebra
(\ref{Algebra}).

We define, for the  algebra of Lourent series (\ref{Algebra}), a trace functional
\begin{equation}
\mbox{tr} f= \int_{-\infty}^\infty u_0\, dx \qquad f\in A,\,\, f=\sum\limits_{-\infty}^\infty u_i(x)p^i
\end{equation}
and a non-degenerate ad-invariant pairing
\begin{equation}
(f,g)= \mbox{tr} (f\cdot g)\qquad f,g\in A.
\end{equation}
Thus we have a Poisson algebra with a commutative multiplication
and unity, the multiplication  satisfies the derivation property
with respect to the Poisson bracket, and  the algebra is equipped
with a non-degenerate ad-invariant pairing. Following \cite{li} we
can define an infinite family of Poisson structures for smooth
functions on the algebra $A$. A function $F$ on $A$ is smooth if
there is a map $dF:A\to A$ such that
$$
F'|_{t=0}(f+tg)=(dF(f),g), \qquad f,g\in A.
$$
The following theorem (\cite{li}, see also \cite{bl}) holds
\begin{theorem}\label{th1}

Let $A$ be a Poisson algebra with commutative multiplication and
unity, Poisson bracket $\{.,.\}$ and non-degenerate, ad-invariant
pairing $(.,.)$. Let the multiplication  satisfies the derivation
property with respect to the Poisson bracket and symmetric with
respect to the pairing, $(fg,h)=(f,gh)$. Assume that $R:A\to A$ is
a classical $r$-matrix.
Then for smooth functions $F$ and $G$ on $A$ \\
a.
\begin{equation}\label{Pst}
\{F,G\}_{(n)}(f)=(f,\{R(f^{n+1}dF(f)),dG(f)\})+(f,\{dF(f),R(f^{n+1}dG(f))\}),
\end{equation}
defines a Poisson structure for each  integer $n\ge -1$.\\
b. The structures $\{.,.\}_{(n)}$ are compatible with each other
(their sum is again a Poisson structure).
\end{theorem}
A linear operator $R:A\to A$ is a classical $r$-matrix  if the bracket
$$
[f,g]=\frac{1}{2}(\{Rf,g\}+\{f,Rg\})
$$
is a Lie bracket.

To apply the above theorem we take an $r$-matrix
\begin{equation}
R=\frac{1}{2}(P_{\ge 0}-P_{\le -1})
\end{equation}
where $P_{\ge 0}$ and $P_{\le -1}$ are projectors on Poisson subalgebras
$$
A_{\ge 0}=\{u=\sum_{0}^\infty u_ip^i:\, u\in A\}\quad \mbox{and}\quad A_{\le 0}=\{u=\sum_{-\infty}^{-1} u_ip^i:\, u\in A\}
$$
respectively. Note that the Lax equation (\ref{Lax_Eqn_Toda1}) is
\begin{equation}
L_t=\{R(L),L\}
\end{equation}
where $L=p+S+Pp^{-1}$.

Using the Poisson structures given by the Theorem (\ref{th1}) we
obtain bi-Hamiltonian formulation of the Toda lattice.

The submanifold $M=\{L\in A:\, L=p+S+Pp^{-1}\}$ is a Poisson
submanifold for the Poisson structure (\ref{Pst}) with $n=-1$.
Restricting this structure on $M$ we obtain the following
Hamiltonian operator
\begin{equation}\label{Ham_Oper_SP-1}
\mathfrak D_{-1}=\left(
\begin{array}{ll}
0 & PD_x+P_x\\
PD_x & 0 \\
\end{array}
\right)
\end{equation}
We have first Hamiltonian formulation for (\ref{Eqn_Toda1})
\begin{equation}
\left(
\begin{array}{l}
S\\
P
\end{array}
\right)_t= \mathfrak D_{-1}\left(
\begin{array}{l}
\delta \mathcal H_{-1}/\delta S\\
\delta \mathcal H_{-1}/\delta P
\end{array}
\right),
\end{equation}
where
\begin{equation}
\mathcal H_{-1}= \frac{1}{2}\mbox{tr} L^2 \quad \mbox{that is}
\quad \mathcal H_{-1}=\frac{1}{2}\int_{-\infty}^\infty (S^2+2P)\,
dx.
\end{equation}
The second Hamiltonian operator can be obtained by restricting the
Poisson structure (\ref{Pst}) with $n=0$ on the submanifold $M$ or
by application of the recursion operator (\ref{Toda_Rec_Oper_SP})
to the Hamiltonian operator (\ref{Ham_Oper_SP-1}). The second
Hamiltonian operator is
\begin{equation}\label{Ham_Oper_SP0}
\mathfrak D_{0}=\left(
\begin{array}{ll}
2PD_x+P_x & SPD_x+SP_x\\
SPD_x+S_xP & P^2D_x+PP_x\\
\end{array}
\right).
\end{equation}
The corresponding Hamiltonian functional is
\begin{equation}
\mathcal H_{0}=\mbox{tr} L \quad \mbox{that is}\quad \mathcal H_{0}=\int_{-\infty}^\infty S\, dx
\end{equation}
Since Hamiltonian operators $\mathfrak {D}_{-1}$ and $\mathfrak
{D}_{0}$ are compatible we have a bi-Hamiltonian representation of
the equation (\ref{Eqn_Toda1}).

In $u,v$ variables (\ref{uv_variables})
the  Hamiltonian operators and functionals take form
\begin{equation}
\mathfrak {B}_{-1}=
\frac{uv}{(u-v)^2}\left(
\begin{array}{cc}
-2u & u+v\\
u+v & -2v\\
\end{array}
\right)D_x+
\end{equation}
\begin{equation}
\frac{1}{(u-v)^3}\left(
\begin{array}{cc}
2uv^2u_x-u^3v_x-u^2vv_x & u^2vv_x+u^3v_x-2uv^2u_x\\
2uv^2v_x-uv^2u_x-v^3u_x & v^3u_x+uv^2u_x-2u^2vv_x\\
\end{array}
\right)
\end{equation}
and
\begin{equation}\label{Ham_uv-1}
\mathcal{G}_{-1}=\int^\infty_{-\infty} (u^2+v^2+4uv)\, dx.
\end{equation}
\begin{equation}
\mathfrak {B}_{0}=\left(
\begin{array}{cc}
0 & uv_x + uvD_x\\
vu_x+ uvD_x & 0\\
\end{array}
\right)
\end{equation}
and
\begin{equation}\label{Ham_uv0}
\mathcal{G}_{0}=\int^\infty_{-\infty} (u+v) \,dx.
\end{equation}
A different approach was used in \cite{fs} to obtain the
Hamiltonian operator $\mathfrak {B}_{0}$ (see also references in
\cite{fs}).
The explicit expressions of an infinite number of
conservation laws for the Toda system(\ref{Eqn_Toda2}) was given
in \cite{fs}
\begin{equation}
Q_{n,t}= F_{n,x}\qquad n=1,2\dots,
\end{equation}
where
\begin{equation}
Q_n=\sum\limits_{j=0}^n \left( n\atop j \right)^2 u^j v^{n-j}\qquad n=1,2,3\dots
\end{equation}
and
\begin{equation}
F_n=\sum\limits_{j=0}^n \frac{n-j}{j+1}\left( n\atop j \right)^2 u^{j+1} v^{n-j} \qquad n=1,2,3\dots .
\end{equation}
The conserved quantities $\mathcal Q_n=\int^\infty_{-\infty} Q_n
dx$ are in involution with respect to the Hamiltonian operators
$\mathfrak{B}_{-1}$ and $\mathfrak{B}_{0}$. One can easily check
if the  boundary conditions preserve the conserved quantities.
\begin{lemma}
For the Toda system(\ref{Eqn_Toda2})with the boundary condition \\
{\bf a.} $(u-v)|_{x=0}=0$ ($(P=\frac{S^2}{4})|_{x=0}$) the
above conservation laws are not preserved;\\
{\bf b.}   $uv|_{x=0}=0$ ($P|_{x=0}=0$)
the quantities
\begin{equation}
\int _0^{\infty} \mathcal Q_n dx \qquad n=1,2,3\dots
\end{equation}
are conserved;\\
{\bf c.} $(u+v)|_{x=0}=0$ ($S|_{x=0}=0$) the quantities
\begin{equation}
\int _0^{\infty} \mathcal Q_n dx \qquad n=2,4,6 \dots
\end{equation}
are conserved.
\end{lemma}

We can use the above Hamiltonian operators to obtain the Hamiltonian
representation of some of the boundary value problems.
\begin{theorem}
The Toda system(\ref{Eqn_Toda2}) on a segment $[0,1]$ with
boundary conditions
\begin{equation}\label{Bound_Cond_01}
uv|_{x=0}=0 \quad \mbox{and}\quad uv|_{x=1}=0.
\end{equation}
admits the bi-Hamiltonian  representation with Hamiltonian
operators ${\mathfrak B}_{(n)}$, n=-1,0, and  Hamiltonians
\begin{equation}
\mathcal {\bar G}_{-1}=\int_0^1 (u^2+v^2+4uv)dx=\int_{-\infty}^\infty (u^2+v^2+4uv)\theta(x)\theta(1-x)dx
\end{equation}
and
\begin{equation}
\mathcal {\bar G}_{0}=\int_0^1 (u+v)dx,=\int_{-\infty}^\infty (u+v)\theta(x)\theta(1-x)dx,
\end{equation}
respectively, where $\theta(x)$ is the Heaviside step function.
\end{theorem}
{\bf Proof.}
The Hamiltonian equations
\begin{equation}
\left(
\begin{array}{l}
u\\
v
\end{array}
\right)_t= \mathfrak{B}_{n}\left(
\begin{array}{l}
\delta \mathcal {\bar G}_{n}/\delta u\\
\delta \mathcal {\bar G}_{n}/\delta v
\end{array}
\right) \quad n=-1,0
\end{equation}
are for  $n=-1$
\begin{equation}
\begin{array}{l}
\displaystyle{u_t=uv_x-\frac{uv}{u-v}(\delta(x)-\delta(1-x)),}\\
\displaystyle{v_t=vu_x+\frac{uv}{u-v}(\delta(x)-\delta(1-x))}
\end{array}
\end{equation}
and for $n=0$
\begin{equation}
\begin{array}{lll}
u_t=uv_x+uv(\delta(x)-\delta(1-x)),\\
v_t=vu_x+uv(\delta(x)-\delta(1-x)),
\end{array}
\end{equation}
where $x\in [0,1]$. Under the boundary conditions $uv|_{x=0}=0$ and $uv|_{x=1}=0$
we have the Toda system(\ref{Eqn_Toda2}) on $[0,1]$. Note that the
Poisson brackets are given by
\begin{equation}
\{\mathcal K,\mathcal N\}=\int_{-\infty}^\infty
\left(\begin{array}{l}
\mathcal \delta \mathcal K/\delta u\\
\mathcal \delta \mathcal K/\delta v\\
\end{array}
\right) {\mathfrak B}_{(n)}
\left(\begin{array}{l}
\mathcal \delta \mathcal N/\delta u\\
\mathcal \delta \mathcal N/\delta v\\
\end{array}
\right)
\end{equation}
where $n=-1,\, 0$.
$\Box$

\section{Integrable boundary conditions for the three field
systems.}\label{sec_3dim_eqn}

Let us  consider a three field hydrodynamic type system on the
algebra (\ref{Algebra}). We take  a Lax function
\begin{equation}\label{Lax_Funct_3dim}
L=p^2+Sp+P+Qp^{-1}.
\end{equation}
We can construct two integrable hierarchies with this Lax function.

The first hierarchy is given by
\begin{equation}\label{Eqn_3dim_hierarchy1}
L_t=\{(L^{n+\frac{1}{2}})_{\ge 0},L\}\qquad n=0,1,2,\dots,
\end{equation}
the first equation of the hierarchy is
\begin{equation}\label{Eqn_3dim_1}
\begin{array}{l}
  S_t=P_x-\frac{1}{2}SS_x,\\
  P_t=Q_x, \\
  Q_t=\frac{1}{2}QS_x.\\
\end{array}
\end{equation}

The second hierarchy is given by
\begin{equation}\label{Eqn_3dim_hierarchy2}
L_t=\{(L^n)_{\ge 0},L\}\qquad n=1,2,3\dots,
\end{equation}
the first equation of the hierarchy is
\begin{equation}\label{Eqn_3dim_2}
\begin{array}{l}
  S_t=2Q_x\\
  P_t=SQ_x+Q S_x\\
  Q_t=QP_x\\
\end{array}
\end{equation}

We also have a recursion operator \cite {gz} of the hierarchies
(\ref{Eqn_3dim_hierarchy1}) and  (\ref{Eqn_3dim_hierarchy2}).
\begin{equation}\label{Rec_Oper_3d}
\left(
\begin{array}{lll}
P-\frac{1}{4}S^2+(\frac{1}{2}P_x-\frac{1}{4}SS_x)D_x^{-1} & \frac{1}{2}S &3+2Q_x D_x^{-1}Q^{-1}\\
\frac{3}{2}Q+\frac{1}{2}Q_xD_x^{-1} & P & 2S+(SQ)_x D_x^{-1}Q^{-1}\\
\frac{1}{4}SQ+\frac{1}{4}S_xQD_x^{-1} & \frac{3}{2}Q & P+ QP_x D_x^{-1}Q^{-1}\\
\end{array} \right)
\end{equation}

The bi-Hamiltonian representation of equations (\ref{Eqn_3dim_1})
and (\ref{Eqn_3dim_2}) is  obtained by restricting the Poisson
structure (\ref{Pst}) with $n=-1$ and $n=0$ on the submanifold
$M=\{L\in A:\, L=p^2+Sp+P+Qp^{-1}\}$. So we have Hamiltonian
operators
\begin{equation}\label{Ham_Oper_3dim-1}
\mathfrak {C}_{-1}=\left(
\begin{array}{lll}
2D_x & 0 & 0 \\
0 & 0 & QD_x+Q_x\\
0 & QD_x & 0\\
\end{array}
\right)
\end{equation}
and
\begin{equation}\label{Ham_Oper_3dim0}
\mathfrak {C}_{0}=\left(
\begin{array}{lll}
(2P-\frac{1}{2}S^2)D_x+P_x-\frac{1}{2}SS_x &3QD_x+2Q_x &\frac{1}{2}QD_x+\frac{1}{2}SQ_x \\
QD_x+Q_x &2SQD_x+SQ_x+QS_x & PQD_x+PD_x\\
\frac{1}{2}SQD_x+\frac{1}{2}QS_x& PD_x+P_xQ &\frac{3}{2}Q^2D_x+ \frac{3}{2}QQ_x \\
\end{array}
\right).
\end{equation}
The equation (\ref{Eqn_3dim_1}) can be written as
\begin{equation}
\left(
\begin{array}{l}
S\\
P\\
Q\\
\end{array}
\right)_t= \mathfrak C_{-1}\left(
\begin{array}{l}
\delta \bar{\mathcal H}_{-1}/\delta S\\
\delta\bar{\mathcal H}_{-1}/\delta P\\
\delta\bar{\mathcal H}_{-1}/\delta Q\\
\end{array}
\right)= \mathfrak C_{0}\left(
\begin{array}{l}
\delta \bar{\mathcal H}_{0}/\delta S\\
\delta \bar{\mathcal H}_{0}/\delta P\\
\delta \bar{\mathcal H}_{0}/\delta Q\\
\end{array}
\right),
\end{equation}
where
\begin{equation}\label{Eqn_3dim_1_Ham_1}
\bar{\mathcal H}_{-1}= \frac{2}{3}\mbox{tr} L^{\frac{3}{2}} \quad \mbox{that is}\quad \bar{\mathcal H}_{-1}=\int_{-\infty}^\infty \left(Q+\frac{1}{2}SP-\frac{1}{24}S^3\right)\, dx
\end{equation}
and
\begin{equation}\label{Eqn_3dim_1_Ham_2}
\bar{\mathcal H}_{0}= 2\mbox{tr} L^{\frac{1}{2}} \quad \mbox{that is}\quad \bar{\mathcal H}_{0}=\int_{-\infty}^\infty S\, dx\, .
\end{equation}

The equation (\ref{Eqn_3dim_2}) can be written as
\begin{equation}
\left(
\begin{array}{l}
S\\
P\\
Q\\
\end{array}
\right)_t= \mathfrak C_{-1}\left(
\begin{array}{l}
\delta \tilde{\mathcal H}_{-1}/\delta S\\
\delta   \tilde{\mathcal H}_{-1}/\delta P\\
\delta  \tilde{\mathcal H}_{-1}/\delta Q\\
\end{array}
\right)= \mathfrak C_{0}\left(
\begin{array}{l}
\delta  \tilde{\mathcal H}_{0}/\delta S\\
\delta  \tilde{\mathcal H}_{0}/\delta P\\
\delta  \tilde{\mathcal H}_{0}/\delta Q\\
\end{array}
\right),
\end{equation}
where
\begin{equation}\label{Eqn_3dim_2_Ham_1}
\tilde{\mathcal H}_{-1}= \frac{1}{2}\mbox{tr} L^{2} \quad \mbox{that is}\quad  \tilde{\mathcal H}_{-1}=\int_{-\infty}^\infty \left(SQ+\frac{1}{2}P^2\right)\, dx
\end{equation}
and
\begin{equation}\label{Eqn_3dim_2_Ham_2}
\tilde{\mathcal H}_{0}= \mbox{tr} L \quad \mbox{that is}\quad  \tilde{\mathcal H}_{0}=\int_{-\infty}^\infty P\, dx\, .
\end{equation}

We can give  both hierarchies in modified variables, writing the
Lax function (\ref{Lax_Funct_3dim}) as $L=p^{-1}(p-u)(p-v)(p-w)$
that is
\begin{equation}
\begin{array}{l}
S=u+v+w,\\
P=uv+uw+vw,\\
  Q=uvw.\\
\end{array}
\end{equation}
To find integrable boundary condition directly for three field
systems is quite difficult. For example, consider hierarchy
(\ref{Eqn_3dim_hierarchy1}). In the following lemmas we use
$P,Q,R$ variables since symmetries and recursion operator have a
simple form in these variables.

\begin{lemma}\label{lemma5}
Let $x=0$ be the boundary. The boundary conditions of the form
$P=F(S)$ and $Q=G(S)$ are compatible with the first symmetry of
the hierarchy (\ref{Eqn_3dim_hierarchy1}) if the functions $F$ and
$G$ satisfy the following differential equations
\begin{equation}\label{lemma5_eq1}
\frac{3}{2}S(F')^2 +3F'G'-\frac{3}{4}F'S^2-3G'S-\frac{3}{2}G=0,
\end{equation}
\begin{equation}\label{lemma5_eq2}
\frac{3}{2}SF'G'+3(G')^2-\frac{3}{2}F'G-\frac{3}{4}G'S^2-\frac{3}{4}SG=0.
\end{equation}
\end{lemma}
{\bf Proof.} The first symmetry of the hierarchy
(\ref{Eqn_3dim_hierarchy1}) is
\begin{equation}
\begin{array}{l}
S_{t_1}=\frac{3}{2}(P-\frac{1}{4}S^2)(P_x-\frac{1}{2}SS_x)+\frac{3}{2}SQ_x+\frac{3}{2}S_xQ,\\
P_{t_1}=\frac{3}{2}PQ_x+\frac{3}{2}P_xQ \frac{3}{4}QSS_x+\frac{3}{8}S^2Q_x,\\
Q_{t_1}=\frac{1}{4}SQ(P_x-\frac{1}{2}SS_x)+\frac{1}{4}Q(P-\frac{1}{4}S^2)+\frac{3}{2}QQ_x+
\frac{1}{2}QPS_x+\frac{1}{2}QSP_x.\\
\end{array}
\end{equation}
Differentiating the boundary conditions $P=F(S),$ $Q=G(S)$ with respect
to the above symmetry and expressing  all the $x$ derivatives in
terms of $t$ derivatives using the equation (\ref{Eqn_3dim_1}) we
obtain the  equations  (\ref{lemma5_eq1}) and (\ref{lemma5_eq2}).
$\Box$

\begin{lemma}\label{lemma6}
Let $x=0$ be the boundary. The boundary condition of the form
$S=F(P,Q)$ is compatible with the first symmetry of the hierarchy
(\ref{Eqn_3dim_hierarchy1}) if function $F$  satisfies the
following differential equations
\begin{multline}\label{lemma6_eq1}
\frac{3}{2}(P-\frac{1}{4}F^2)F_P+\frac{3}{2}F=\\
\frac{3}{2}PF_P+\frac{3}{2}QF^2_P+\frac{3}{8}QF^2_P+\frac{3}{8}F^2F_P+
\frac{1}{4}QFF_Q+\frac{3}{2}QF_Q+\frac{1}{2}FF_PF_Q,
\end{multline}
\begin{multline}\label{lemma6_eq2}
\frac{3}{2}(P-\frac{1}{4}F^2)F_Q+\frac{3}{2}=\\
\frac{3}{2}QF_PF_Q+\frac{3}{2}FF_P+
\frac{1}{2}(P-\frac{1}{4}S^2)F_Q+PF_Q+
\frac{1}{2}QFF^2_Q+\frac{1}{2}F^2F_Q.
\end{multline}
\end{lemma}
{\bf Proof.} We differentiate the boundary condition $S=F(P,Q)$
with respect to the symmetry (\ref{Eqn_3dim_hierarchy1}) and
express  all the $x$ derivatives in terms of $t$ derivatives using
the equation (\ref{Eqn_3dim_1}). Then   separating terms
containing $P_t$ and $Q_t$ we obtain the  equations
(\ref{lemma6_eq1}) and (\ref{lemma6_eq2}). $\Box$

\bigskip

\noindent The differential equations obtained in the above lemmas
are nonlinear partial differential equations which are rather
complicated. So, to obtain integrable boundary conditions it is
easy to use  integrable reductions discussed in section
\ref{Sec_Integr_Red}. Let  $x=0$ be a boundary.

\noindent
{\bf a.} Integrable reduction $u=v$ gives integrable
boundary condition $u|_{x=0}=v|_{x=0}$  or
$(S^3Q-S^2P^2+4Q^3-18SPQ+27Q^2)|_{x=0}=0$ (condition on
coefficients of cubic equation to have at least to equal roots) in
$S$, $P$, $Q$ variables.

\noindent {\bf b.} Integrable reduction $u=v=w$ gives integrable
boundary conditions $u|_{x=0}=v|_{x=0}=w|_{x=0}$ or
$P|_{x=0}=\frac{1}{3}S^2|_{x=0}$, $Q|_{x=0}=\frac{1}{27}S^3|_{x=0}$ (condition on coefficients
of cubic equation to have all roots equal.

\noindent
{\bf c.}
Integrable reduction $u=0$ gives integrable boundary
condition $u|_{x=0}=0$ or $Q|_{x=0}=0$ .

\noindent
{\bf d.}
Integrable reduction $u=0$, $v=0$ gives integrable boundary
conditions $u|_{x=0}=0$, $v|_{x=0}=0$ or $P|_{x=0}=0$, $Q|_{x=0}=0$.
To obtain  boundary value problems that admit bi-Hamiltonian
representation we modify  Hamiltonian functions, as in the case of
Toda system. We use $S,P,Q$ variables, the Hamiltonian operators
have simpler form in this variables.

For equation (\ref{Eqn_3dim_1}) we have
\begin{theorem}
The  equation (\ref{Eqn_3dim_1}) on a segment $[0,1]$
with boundary conditions
\begin{equation}\label{Eqn_3dim_1_Bound_Cond}
\left(P-\frac{1}{4}S^2\right)|_{x=0}=0,\, Q|_{x=0}=0 \quad \mbox{and}\quad \left(P-\frac{1}{4}S^2\right)|_{x=1}=0,\, Q|_{x=1}=0.
\end{equation}
admits the bi-Hamiltonian  representation with Hamiltonian
operators (\ref{Ham_Oper_3dim-1}), (\ref{Ham_Oper_3dim0}) and  Hamiltonians
\begin{equation}
\bar{\bar {\mathcal H}}_{-1}=\int_{-\infty}^\infty \left(Q+\frac{1}{2}SP-\frac{1}{24}S^3\right)\theta(x)\theta(1-x)dx
\end{equation}
and
\begin{equation}
\bar{\bar{\mathcal H}}_{0}=\int_{-\infty}^\infty S\theta(x)\theta(1-x)dx,
\end{equation}
respectively, where $\theta(x)$ is the Heaviside step function.
\end{theorem}
{\bf Proof.}
The Hamiltonian equations
\begin{equation}
\left(
\begin{array}{l}
S\\
P\\
Q\\
\end{array}
\right)_t= \mathfrak{C}_{n}\left(
\begin{array}{l}
\delta \bar{\bar{\mathcal H}}_{n}/\delta S\\
\delta \bar{\bar{\mathcal H}}_{n}/\delta P\\
\delta \bar{\bar{\mathcal H}}_{n}/\delta Q\\
\end{array}
\right) \quad n=-1,0
\end{equation}
are for  $n=-1$
\begin{equation}
\begin{array}{l}
S_t=P_x-\frac{1}{2}SS_x+(P-\frac{1}{4}S^2)(\delta(x)-\delta(1-x)),\\
  P_t=Q_x +\frac{1}{2}SQ(\delta(x)-\delta(1-x)),\\
  Q_t=\frac{1}{2}QS_x+Q(\delta(x)-\delta(1-x))\\
\end{array}
\end{equation}
and for $n=0$
\begin{equation}
\begin{array}{lll}
S_t=P_x-\frac{1}{2}SS_x+(2P-\frac{1}{2}S^2)(\delta(x)-\delta(1-x)),\\
  P_t=Q_x +Q(\delta(x)-\delta(1-x)),\\
  Q_t=\frac{1}{2}QS_x+\frac{1}{2}SQ(\delta(x)-\delta(1-x)),\\
\end{array}
\end{equation}
where $x\in [0,1]$. Under the boundary conditions (\ref{Eqn_3dim_1_Bound_Cond})
we have the  equation (\ref{Eqn_3dim_1}) on $[0,1]$.
$\Box$

\noindent  The boundary conditions
(\ref{Eqn_3dim_1_Bound_Cond}) are symmetry integrable.

\begin{lemma}\label{lemma8}
All the symmetries of the hierarchy (\ref{Eqn_3dim_hierarchy1}) are
compatible with the boundary condition (\ref{Eqn_3dim_1_Bound_Cond}).
\end{lemma}
{\bf Proof.}
The boundary condition (\ref{Eqn_3dim_1_Bound_Cond}) is compatible with an evolution symmetry
\begin{equation}\label{s6l8e1}
\left(
\begin{array}{l}
S\\
P\\
Q
\end{array}
\right)_\tau=
\left(
\begin{array}{l}
\sigma\\
\pi\\
\kappa
\end{array}
\right)
\end{equation}
if $\pi=\frac{1}{2}S\sigma$ and $\kappa=0$ for $P=\frac{1}{4}S^2$ and $Q=0$ on the boundary $x=0$. That is under the conditions (\ref{Eqn_3dim_1_Bound_Cond}) the   symmetry (\ref{s6l8e1}) should take the form
\begin{equation}\label{s6l8e2}
\left(
\begin{array}{l}
S_\tau\\
\frac{1}{2}SS_\tau\\
0
\end{array}
\right)=
\left(
\begin{array}{l}
\sigma\\
\frac{1}{2}S\sigma\\
0
\end{array}
\right)
\end{equation}
One can check that the first symmetry of the hierarchy (\ref{Eqn_3dim_hierarchy1})
has such a form. Let us show that the recursion operator (\ref {Rec_Oper_3d})
  preserves the form (\ref{s6l8e2}).
On the boundary $x=0$ we rewrite the recursion operator (\ref {Rec_Oper_3d})
in terms of $t$ derivatives
using the equation (\ref{Eqn_3dim_1}) as follows
\begin{equation}\label{Rec_Oper_3d_syst1t}
\left(
\begin{array}{lll}
P-\frac{1}{4}S^2-\frac{1}{4}S_tD_t^{-1}S & \frac{1}{2}S+ \frac{1}{2}S_tD_t^{-1}&3+P_t D_t^{-1}\\
\frac{3}{2}Q-\frac{1}{4}P_tD_t^{-1}S & P+ \frac{1}{2}P_tD_t^{-1}& 2S+(\frac{1}{2}SP_t+Q_t)D_t^{-1}\\
\frac{1}{4}SQ-\frac{1}{4}Q_tD_t^{-1}S & \frac{3}{2}Q+ \frac{1}{2}Q_tD_t^{-1}& P+ \frac{1}{2}QP_t D_t^{-1}\\
\end{array} \right)
\end{equation}
Applying the recursion operator (\ref{Rec_Oper_3d_syst1t}) to a symmetry (\ref{s6l8e2}) we obtaine a symmetry
\begin{equation}
\left(
\begin{array}{l}
S_{\tilde\tau}\\
\frac{1}{2}SS_{\tilde\tau}\\
0
\end{array}
\right)=
\left(
\begin{array}{l}
\tilde\sigma\\
\frac{1}{2}S\tilde\sigma\\
0
\end{array}
\right).
\end{equation}
$\Box$

\noindent
For equation (\ref{Eqn_3dim_2}) we have
\begin{theorem}
The  equation (\ref{Eqn_3dim_2}) on a segment $[0,1]$
with boundary conditions
\begin{equation}\label{Eqn_3dim_2_Bound_Cond}
Q|_{x=0}=0 \quad \mbox{and}\quad  Q|_{x=1}=0.
\end{equation}
admits the bi-Hamiltonian  representation with Hamiltonian
operators (\ref{Ham_Oper_3dim0}), (\ref{Ham_Oper_3dim0}) and  Hamiltonians
\begin{equation}
\tilde{\tilde{\mathcal H}}_{-1}=\int_{-\infty}^\infty \left(SQ+\frac{1}{2}P^2\right)\theta(x)\theta(1-x)dx
\end{equation}
and
\begin{equation}
\tilde{\tilde{\mathcal H}}_{0}=\int_{-\infty}^\infty P\theta(x)\theta(1-x)dx,
\end{equation}
respectively, where $\theta(x)$ is the Heaviside step function.
\end{theorem}
{\bf Proof.}
The Hamiltonian equations
\begin{equation}
\left(
\begin{array}{l}
S\\
P\\
Q\\
\end{array}
\right)_t= \mathfrak{C}_{n}\left(
\begin{array}{l}
\delta \tilde{\tilde{\mathcal H}}_{n}/\delta S\\
\delta \tilde{\tilde{\mathcal H}}_{n}/\delta P\\
\delta \tilde{\tilde{\mathcal H}}_{n}/\delta Q\\
\end{array}
\right) \quad n=-1,0
\end{equation}
are for  $n=-1$
\begin{equation}
\begin{array}{l}
S_t=2Q_x+2Q(\delta(x)-\delta(1-x)),\\
  P_t=SQ_x +QS_x+PQ(\delta(x)-\delta(1-x)),\\
  Q_t=QP_x+SQ(\delta(x)-\delta(1-x))\\
\end{array}
\end{equation}
and for $n=0$
\begin{equation}
\begin{array}{lll}
S_t=2Q_x+Q(\delta(x)-\delta(1-x)),\\
  P_t=SQ_x +QS_x+2SQ(\delta(x)-\delta(1-x),\\
  Q_t=QP_x+PQ(\delta(x)-\delta(1-x)),\\
\end{array}
\end{equation}
where $x\in [0,1]$. Under the boundary conditions (\ref{Eqn_3dim_2_Bound_Cond})
we have the  equation (\ref{Eqn_3dim_2}) on $[0,1]$.
$\Box$

\noindent In the same way as in lemma \ref{lemma8} one can show
that the boundary condition (\ref{Eqn_3dim_2_Bound_Cond}) is
symmetry integrable. This case is similar to the case of Toda
system (the boundary condition $Q|_\Gamma=0$ in modified variables
is $uvw|_\Gamma=0$).

\section {Conclusion}

In this article we studied the problem of integrable boundary
conditions for hydrodynamic type integrable systems. To our
knowledge the problem has never been discussed in the literature
before. Since the term integrability has various meanings the
notion of integrable boundary conditions has also several
definitions. As basic ones we take three definitions, namely,
consistency with infinite set of symmetries, consistency with
infinite set of conserved quantities, and consistency with the
Hamiltonian integrability (or bi-Hamiltonian structure).
Comparison of these three kinds of integrable boundary conditions
shows that the consistency with the bi-Hamiltonian structure is a
very severe restriction. Only very special kind of boundary
conditions passes this test. The class of symmetry consistent
boundary conditions seems to be relatively larger. As an example
we studied dispersionless Toda system as an example of $N=2$
system. We found all symmetry compatible boundary conditions of
this system and showed that only a subclass of these boundary
conditions are compatible with the Hamiltonian formulation of the
system. We pointed out that the integrable reductions of the $N-$
system of hydrodynamical type of equations are directly related to
the integrable boundary conditions of the same systems. Using this
property,  a method for constructing exact solutions
satisfying the integrable boundary conditions is given. We considered also
an $N=3$ system.  Integrable boundary conditions
compatible with symmetries and compatible with the
Hamiltonian formulation of this system were found. 
\section{Acknowledgement}

This work is partially supported by the Scientific and
Technological Research Council of Turkey (TUBITAK) and  Turkish
Academy of Sciences (TUBA). One of the authors (I.H.) thanks also RFBR grant $\sharp$ 06-01-92051 KE-a.

\end{document}